\documentclass[twocolumn,showpacs,preprintnumbers,amsmath,amssymb]{revtex4}

\usepackage{graphicx}
\usepackage{dcolumn}
\usepackage{bm}

\newcommand{\pabl}[2]{\frac{\partial #1}{\partial #2}}
\newcommand{\imai}{{\rm i}}
\newcommand{\e}{{\rm e}}
\renewcommand{\d}{{\rm d}}
\begin{document}

\preprint{To be published in Physical Review B (2002)}

\title{Gain in quantum cascade lasers and superlattices:
A quantum transport theory}

\author{Andreas Wacker}
\email{wacker@physik.tu-berlin.de}
\affiliation{Institut f{\"u}r Theoretische Physik, 
Technische Universit{\"a}t Berlin, Hardenbergstr.~36, 10623~Berlin, Germany}
\date{13. June 2002}

\begin{abstract}
Gain in current-driven  semiconductor heterostructure devices
is calculated within the theory of  
nonequilibrium Green functions.
In order to treat the nonequilibrium distribution self-consistently
the full two-time structure of the theory is employed
without relying on any sort of Kadanoff-Baym Ansatz.
The results are independent of the choice of the
electromagnetic field if the variation of the self-energy is
taken into account.
Excellent quantitative agreement is obtained with the experimental 
gain spectrum of a quantum cascade laser. Calculations for
semiconductor superlattices show that 
the simple 2-time miniband transport model gives reliable results
for large miniband widths at room temperature.

\end{abstract}

\maketitle
 
\section{Introduction}
The prospect of a semiconductor laser in the infrared
and THz region has been one of the key reasons for 
the development and study of semiconductor heterostructure
elements, since the first proposal of semiconductor superlattices
in 1970 \cite{ESA70}.
A possible gain mechanism may be based on two
different ideas:
(i) At certain electrical fields resonant tunneling between
different subbands can lead to population inversion
associated with gain at the transition energy \cite{KAZ71}. 
This idea was realized in the quantum cascade laser \cite{FAI94a},
which has become an important device in the infrared region.
Lasing in the THz region has been demonstrated
very recently as well \cite{KOE02}.
For further details, see the review \cite{GMA01}.
(ii) The occurrence of negative differential
conductivity in superlattices gives raise
to gain in the low frequency range extending up to
frequencies of the order of the Bloch frequency \cite{KTI72}.
Despite a strong effort by several groups all over the world, 
this idea is still not realized. The difficulty in its realization
is attributed to the instability of the operating state
leading to domain formation \cite{REN98}, which may be circumvented
by different strategies \cite{WAC97e,KRO00,ALL01}.
For further references and a detailed discussion
see Ref.~\onlinecite{WAC02}.

These concepts for gain are commonly described in different ways:
(i) An external electromagnetic field
causes transitions between different levels, where
the transition rate $R$ is evaluated by Fermi's golden rule.
If stimulated emission dominates stimulated absorption
(i.e., for population inversion), gain occurs at the transition
frequency. For a pair of states $\alpha,\beta$ with
energies $E_{\alpha},E_{\beta}$, occupation probabilities
$f_{\alpha},f_{\beta}$, and a dipole matrix element
$z_{\alpha\beta}$, the
contribution to the material gain 
(i.e., increase of light intensity per length) 
is given by
\begin{equation}\begin{split}
G_m(\omega)=&\frac{1}{V}
\frac{R^{\text{stim. em.}}_{\alpha \to \beta}
-R^{\text{stim. abs.}}_{\beta \to \alpha}}{\mbox{Photon flux per area}}\\
=& \frac{\pi\omega\left|ez_{\alpha\beta}\right|^2}
{\sqrt{\epsilon_r}c\epsilon_0V}
\delta(E_{\alpha}-E_{\beta}-\hbar\omega)(f_{\alpha}-f_{\beta})\, ,
\label{EqGainSimple}
\end{split}\end{equation}  
where $V$ is the normalization volume,  $e<0$ is the electron charge,
$c$ is the vacuum speed of light, and $\epsilon_r$ 
the background dielectric constant. 

(ii) Transport theory in the presence of alternating electric fields
provides the complex 
dynamical conductance $\sigma(\omega)$.
Standard electrodynamics (see, e.g., section 7.5 of Ref.~\cite{JAC98a})
gives the material gain 
(the negative absorption coefficient $\alpha$, which is assumed to
be small compared to the wave vector here)
\begin{equation}
G_m(\omega)\approx -\frac{\Re\{\sigma(\omega)\}}
{c\epsilon_0\sqrt{\epsilon_r-\Im\{\sigma(\omega)\}/\epsilon_0\omega}}
\, .
\label{EqGainSigma}
\end{equation}
Thus, gain is equivalent to a negative differential conductivity
in the respective frequency range. 
This concept has been frequently applied to superlattice transport.
(For superlattices Eq.~(\ref{EqGainSimple}) gives zero gain 
in the basis of Wannier Stark states, as the translational
symmetry gives identical $f_{\alpha}$ for all Wannier-Stark levels.)
Here the current is either evaluated from a semiclassical
miniband transport model \cite{KTI72,REN98,KRO00,ALL01}
or by sequential tunneling between  
different layers \cite{TUC85,WAC97e,PLA97}. While these approaches use
a macroscopic current density in Eq.~(\ref{EqGainSigma}), the
consideration of polarization currents allows for a derivation of
Eq.~(\ref{EqGainSimple}), see, e.g., Ref.~\onlinecite{HAU94}. 

For complicated semiconductor heterostructures, with a
variety of different tunneling and optical transitions,
gain may result both from optical transitions in the spirit
of (i) and macroscopic currents in the spirit of (ii).
The aim of this paper is to show the
feasibility of a general approach
based on a quantum transport theory
which treats both mechanisms on a equal footing.

The method of nonequilibrium Green functions \cite{KAD62}
is used here, which has been frequently  applied to light emission
in semiconductors 
(see, e.g., Refs.~\onlinecite{SCH88h,HAU96,HEN96,PER98}
and references given therein).
While these works consider transitions between
the conduction and valence band, the focus in this paper is on
transitions within the conduction band in current-driven
heterostructure devices. 
The calculations are performed in two steps:
First the transport problem is solved by evaluating
a self-consistent stationary solution of the quantum kinetic equations 
for a given applied bias.
This provides us with the Green functions
describing the electronic state far from equilibrium.
In a second step, an additional weak radiation field is taken into
account. The time-dependent kinetic equations are linearized
around the stationary nonequilibrium state in oder to 
study the linear response of the current-driven system.
The formulation used here employs the full two-time structure
of the theory without any sort of (generalized) Kadanoff-Baym ansatz
\cite{LIP86}.
The capability of the approach is demonstrated by
calculations of the gain spectra in a quantum cascade laser
and a superlattice.

\section{Theory}
We describe a general system by a set of orthonormal states
labeled $\alpha$ with energies $E_{\alpha}$ and write the
Hamiltonian in the form:
\begin{equation}
\hat{H}=
\sum_{\alpha}E_{\alpha}\hat{a}^{\dag}_{\alpha}\hat{a}_{\alpha}
+\sum_{\alpha\beta}U_{\alpha\beta}(t)\hat{a}^{\dag}_{\alpha}\hat{a}_{\beta}
+\hat{H}_{\textrm{scatt}}
\end{equation}
where $\hat{a}_{\alpha}$ and $\hat{a}^{\dag}_{\alpha}$  are
electron annihilation and creation operators in the state $\alpha$.
$U_{\alpha\beta}(t)$ represents the
matrix elements of the
kinetic energy and the potential part of the Hamiltonian $\hat{H}$.
In the following, $U(t)$ will be split into a constant part
$\tilde{U}$ and a small time-dependent perturbation
$\delta{U}(t)$ describing the interaction with a radiation field.
Finally, $\hat{H}_{\textrm{scatt}}$ contains the part of the Hamiltonian
which will be solved perturbatively (such as impurity, phonon, or
electron-electron scattering matrix elements).

Within the theory of nonequilibrium Green functions \cite{KAD62,HAU96}
the key quantities are
the correlation function (or 'lesser' Green function)
\begin{equation}\begin{split}
G^<_{\alpha_1,\alpha_2}(t_1,t_2)=
\imai\Big\langle \hat{a}^{\dag}_{\alpha_2}(t_2)
\hat{a}_{\alpha_1}(t_1)\Big\rangle
\end{split}\end{equation}
and the retarded/advanced Green functions
\begin{equation}\begin{split}
G^{\text{ret/adv}}_{\alpha_1,\alpha_2}&(t_1,t_2)= 
\mp \imai\Theta[\pm (t_1-t_2)]\\
&\times \Big\langle 
\hat{a}_{\alpha_1}(t_1)\hat{a}^{\dag}_{\alpha_2}(t_2)
+\hat{a}^{\dag}_{\alpha_2}(t_2)\hat{a}_{\alpha_1}(t_1)
\Big\rangle\, ,
\end{split}\end{equation}
respectively, where the Heisenberg picture is used.

First consider a stationary state with a time independent
matrix $U(t)=\tilde{U}$, neglecting the radiation field.
In this case all functions depend only on the time 
difference $t_1-t_2$ [these
stationary state functions are labeled by a tilde, e.g., 
$\tilde{G}(t_1-t_2)$, in the following],
and it is convenient to work in Fourier space defined by
\begin{equation}
\tilde{G}_{\alpha_1,\alpha_2}(E)=\frac{1}{\hbar}\int \d t\,
\e^{\imai Et/\hbar} \tilde{G}_{\alpha_1,\alpha_2}(t)\, .
\label{EqFourier-tilde}
\end{equation}
Then Eqs.~(\ref{EqGret-time-tilde},\ref{EqKeldysh-time-tilde}) of the appendix
become the matrix equations
\begin{gather}
\begin{split} 
\left(E-E_{\alpha_1}\right) 
\tilde{G}^{\rm ret}_{\alpha_1,\alpha_2}(E)
-\sum_{\beta} \tilde{U}_{\alpha_1,\beta}\
\tilde{G}^{\rm ret}_{\beta,\alpha_2}(E) 
\\ 
=\delta_{\alpha_1,\alpha_2}+ 
\sum_{\beta}\tilde{\Sigma}^{\rm ret}_{\alpha_1,\beta}(E) 
\tilde{G}^{\rm ret}_{\beta,\alpha_2}(E) \label{EqDysonE-tilde}
\end{split}\\
\tilde{G}^<_{\alpha_1,\alpha_2}(E)=\sum_{\beta,\beta'} 
\tilde{G}^{\rm ret}_{\alpha_1,\beta}(E)\tilde{\Sigma}^{<}_{\beta,\beta'}(E) 
\tilde{G}^{\rm adv}_{\beta',\alpha_2}(E)\label{EqKeldyshE-tilde}\, . 
\end{gather}
These have to be solved
self-consistently together with the equations for the 
self-energies which are
functionals of the Green functions
\begin{equation}
\tilde{\Sigma}(E)={\cal F}_{E}
\left\{\tilde{G}^{\textrm{ret}}(E'),\tilde{G}^{<}(E')\right\}\, .
\label{EqFuncSigmaE}
\end{equation}
This standard approach allows for a self-consistent evaluation
of the Green functions $\tilde{G}$ in a nonequilibrium situation
caused by an applied bias (contained in $\tilde{U}$).
Details, such as the specific form of the functionals
in the self-consistent Born approximation,
can be found in, e.g., \cite{DAT95,LAK97,WAC02}.

The current density (in the $z$-direction) 
can be evaluated from the expectation value
of the momentum operator divided by the mass:
\begin{equation}\begin{split}
J_z=&\frac{e}{V} 
\langle \frac{\hat{p}_z}{m}\rangle
=\frac{e}{V} 
\frac{\imai}{\hbar}\langle [\hat{H},\hat{z}]\rangle\\
=&\frac{e}{\hbar V}\sum_{\alpha\beta}W_{\alpha\beta}
G^<_{\beta\alpha}(t,t)+\mbox{Scattering currents}
\label{EqCurrent}
\end{split}\end{equation}
where $W_{\alpha\beta}=\sum_{\gamma}
(U_{\alpha\gamma}z_{\gamma\beta}-z_{\alpha\gamma}U_{\gamma\beta})$. 
The scattering currents result
from the scattering part of the Hamiltonian and can be expressed
in terms of self-energies. Details will be given elsewhere.               

Now we consider the influence of an additional time-dependent
potential 
\begin{equation}
\delta {\bf U}(t)=\int \frac{\d \omega}{2\pi} \delta {\bf U}(\omega)
\e^{-\imai\omega t}
\end{equation}
in the Hamiltonian.
(Bold capital symbols denote matrices in the state indices
$\alpha,\beta$). 
We treat the change of the system in linear response,
and set
\begin{equation}
G(t_1,t_2)=\tilde{G}(t_1-t_2)+ 
\delta G(t_1,t_2)\, .
\end{equation}
The chance in the self-energies 
is described within the linearization of the functional
(here in the time domain)
\begin{equation}
\Sigma(t_1,t_2)={\cal F}_{t} \left\{\tilde{G}+\delta G\right\}
\approx\tilde{\Sigma}(t_1-t_2)+\delta\Sigma(t_1,t_2)\, .
\label{EqFuncSigmaTime}
\end{equation}
Such a decomposition has been used in Ref.~\onlinecite{SCH88h} for
Hartree-Fock self-energies, e.g.. In this case, the time-dependence of
the self-energies allows for a reduction to density matrix equations
by setting $t_1=t_2$.
In contrast, in our case, where scattering effects are considered, 
$\delta G$ exhibits an explicit time dependence 
in both arguments.
This two time structure is fully taken into account here.
Therefore we apply the Fourier decomposition in both times via
\begin{equation}
\delta {\bf G}(t_1,t_2)=
\int \frac{\d \omega}{2\pi} \e^{-\imai\omega t_1}
\int\frac{\d E}{2\pi} \delta {\bf G}(\omega,E)  \e^{-\imai E(t_1-t_2)/\hbar} 
\label{EqFourier}
\end{equation}
The same decomposition is used for $\delta \Sigma$.
It is shown in appendix \ref{AppDer} that 
$\delta G$ is determined within linear response 
by the following equations:
\begin{widetext}
\begin{eqnarray}
\delta {\bf G}^{\text{ret/adv}}(\omega,E)&=&
\tilde{\bf G}^{\text{ret/adv}}(E+\hbar \omega)
\left[\delta {\bf U}(\omega)+\delta\bm{\Sigma}^{\text{ret/adv}}
(\omega,E)\right]
\tilde{\bf G}^{\text{ret/adv}}(E)
\label{EqGretLinResp}\\
\delta {\bf G}^{<}(\omega,E)&=&
\tilde{\bf G}^{\text{ret}}(E+\hbar \omega)\delta {\bf U}(\omega)
\tilde{\bf G}^{<}(E)
+\tilde{\bf G}^{<}(E+\hbar \omega)\delta{\bf U}(\omega)
\tilde{\bf G}^{\text{adv}}(E)\nonumber \\
&+&\tilde{\bf G}^{\text{ret}}(E+\hbar\omega)
\delta\bm{\Sigma}^{\text{ret}}(\omega,E)\tilde{\bf G}^{<}(E)
+\tilde{\bf G}^{\text{ret}}(E+\hbar\omega)
\delta\bm{\Sigma}^{<}(\omega,E)\tilde{\bf G}^{\text{adv}}(E)\nonumber \\
&+&\tilde{\bf G}^{<}(E+\hbar \omega)
\delta\bm{\Sigma}^{\text{adv}}(\omega,E)\tilde{\bf G}^{\text{adv}}(E)
\label{EqGlessLinResp}
\end{eqnarray}
\end{widetext}

The changes in self-energy $\delta \Sigma$ are functionals of 
$\delta G$, which 
have to be evaluated self-consistently with $\delta G$.
The derivation of the functionals from the linearization
of Eq.~(\ref{EqFuncSigmaTime}) is straightforward and can be
performed for arbitrary self-energies. The situation
is particularly simple for the self-consistent Born approximation,
where the functional ${\cal F}$ is linear in $G$, and
one finds: $\delta \Sigma(\omega,E)={\cal F}_E
\left\{\delta G^{\textrm{ret}}(\omega,E')
,\delta G^{<}(\omega,E')\right\}$
with the same functional as Eq.~(\ref{EqFuncSigmaE}).
(Note, that $\delta {\bf G}^{\mathrm{adv}}(\omega,E')
=\left[\delta {\bf G}^{\mathrm{ret}}(\omega,E')\right]^{\dag}$ does
{\em not hold} for the time-dependent quantities. Thus, the advanced
quantities should be calculated explicitely)

The terms related to $\delta \bm{\Sigma}$  correspond to the
ladder corrections in the evaluation of diagrams for
the Kubo formula in equilibrium. 
It is a general advantage of linear response within nonequilibrium 
Green functions that these terms are obtained directly,
see also \cite{HAU96,WAC99}.

Now we consider the linear response to an external radiation field
where the electric field points in the $z$ direction.
This gives additional terms in the Hamiltonian
$\hat{H}=(\hat{\vec{p}}-e\vec{A})^2/2m +V(\vec{r})+e\varphi(\vec{r})$,
where $\vec{A},\varphi$ are the electromagnetic potentials.
Neglecting quadratic terms in the radiation field
and terms containing $k$ (i.e., assuming that the wavelength 
is large compared to the size of the active region),
the following perturbation potentials are obtained (see Appendix
\ref{AppGauge}):
In the {\em Coulomb gauge}
\begin{equation}
\delta U_{\alpha,\beta}(\omega)=
\frac{e}{\imai\omega}F(\omega)
\left(\frac{\hat{p}_z}{m}\right)_{\alpha,\beta}
\approx -\frac{eF(\omega)}{\hbar \omega}
W_{\alpha,\beta}
\label{EqDeltaU-Coulomb}
\end{equation}         
neglecting the scattering part of $\hat{H}$; in the {\em Lorentz gauge}
\begin{equation}
\delta U_{\alpha,\beta}(\omega)=
-eF(\omega) z_{\alpha,\beta}\, .
\label{EqDeltaU-Lorentz}
\end{equation}      

Finally, the change in current density is
given by the change in Eq.~(\ref{EqCurrent}) 
\begin{equation}
\begin{split}
\delta J_z(\omega)
=\frac{e}{\hbar V}
\int \frac{\d E}{2\pi}
{\rm Tr}\Big\{ &
\left[\delta {\bf U}(\omega)\cdot {\bf Z}-
{\bf Z}\cdot \delta {\bf U}(\omega)\right]\\
&\cdot\tilde{\bf G}^<(E)+{\bf W}\cdot\delta {\bf G}^<(\omega, E)
\Big\}\, .
\end{split}\end{equation}
The scattering current in Eq.~(\ref{EqCurrent}) 
has not been included here.
With these ingredients, one obtains the complex
conductivity $\sigma(\omega)=\delta J(\omega)/F(\omega)$, as
well as the gain coefficient via Eq.~(\ref{EqGainSigma}).

\section{Results}

In the following we consider two different structures:
(i) The quantum-cascade-laser structure of Ref.~\onlinecite{SIR98}.
(ii) The superlattice used in Ref.~\onlinecite{ALL01} with
2 nm Al$_{0.3}$Ga$_{0.7}$As barriers and 8 nm GaAs wells and a doping of 
$n=10^{16}/{\rm cm}^3$.
In both cases, we use a basis set consisting of products
of Wannier functions (in the growth direction $z$)
and plane waves with wave vector ${\bf k}$
(in the $(x,y)$-plane), and
obtain $U_{\alpha\beta}$ 
from nominal sample parameters \cite{WAC02}.
The self-consistent solution for the electrostatic potential
is included as well.

\begin{figure}
\includegraphics[width=0.9\columnwidth]{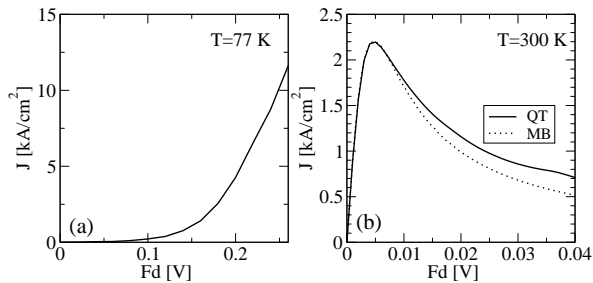}
\caption[a]{Current density versus potential drop per period
for the  quantum cascade laser of Ref.~\onlinecite{SIR98} 
(a) and
the superlattice structure of Ref.~\onlinecite{ALL01} (b).
The full line gives the result of the quantum transport model (QT),
while the dashed line is evaluated with the simple 2-time miniband model (MB),
Eq.~(\ref{EqMiniCurrent}).}
\label{FigKenn}
\end{figure}

For the scattering Hamiltonian we include interface-roughness scattering
and phonon scattering (including both optical phonons and
a second phonon with low energy to mimic acoustic phonons).
The self-energies are evaluated in the self-consistent 
Born approximation. Furthermore, we use momentum-independent scattering 
matrix elements (evaluated for a typical momentum transfer) 
and diagonal self-energies. In this approximation the
self-energies are not ${\bf k}$-dependent, which
significantly reduces the numerical effort.
Results and further details for superlattices \cite{WAC99b,WAC02} 
and quantum cascade lasers \cite{LEE02} have been given previously.
The current-field characteristics are displayed in
Fig.~\ref{FigKenn}(a,b) for both structures, which agree reasonably
well with the respective experimental data.
Note that domain formation in the region of
negative differential conductivity modifies the behavior
of the experimental curve for the superlattice. Such effects
are beyond the scope of the general approach discussed here,
but can be easily studied using simpler transport models 
\cite{BON95,PAT98,WAC02}.
Similar results for the current-field relation of
this quantum cascade laser have been reported in Ref.~\onlinecite{IOT01a}.

For the superlattice structure, the simple
2-time miniband transport model \cite{KTI72,SHI75}
gives 
\begin{equation}
J_{\mathrm MB}(F)=en\frac{d\Delta}{2\hbar}
\frac{I_1(\Delta/2k_BT)}{I_0(\Delta/2k_BT)}\,
\frac{eFd\, \hbar/\tau_e}{(eFd)^2+\hbar^2/\tau_e\tau_m}
\label{EqMiniCurrent}
\end{equation}
for a non-degenerate electron gas. Here $F$ is the
electrical field, $\Delta$ the miniband width and
$d$ the period of the superlattice. 
$\tau_e=0.17$ ps and $\tau_m=0.113$ ps represent fitted, 
phenomenological  energy and momentum scattering times.
$I_j(x)$ are the modified Bessel functions.
In Fig.~\ref{FigKenn}(b) this model shows 
reasonable agreement with the full quantum transport
calculation for $|eFd|\lesssim \Delta/2=11$ meV, the field
range where miniband transport holds true 
\cite{WAC98a}. (For lower temperatures
the shape of the current-field relation becomes more complicated
both  in the semiclassical Boltzmann and the quantum transport
model \cite{WAC99b} so that the simple 2-time model cannot fit the data.)

\begin{figure}
\includegraphics[width=0.9\columnwidth]{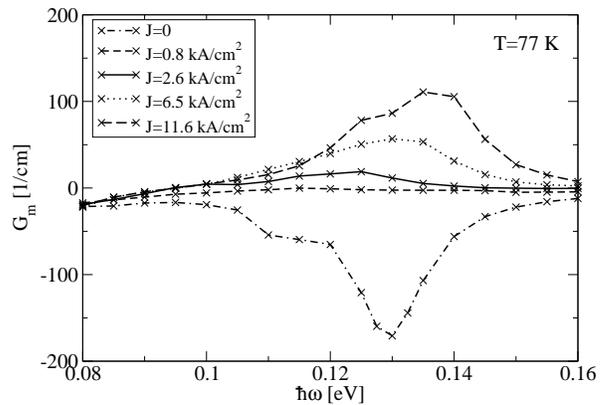}
\caption{Gain spectrum for the quantum cascade laser
of Ref.~\onlinecite{SIR98}
evaluated from the quantum transport model
with Coulomb gauge and neglecting self-energy corrections.}
\label{FigGainQCL}
\end{figure}

In the following the response to an external radiation field
is studied:
First let us use the Coulomb gauge and neglect the terms with 
$\delta \Sigma$ in Eq.~(\ref{EqGlessLinResp}).
The gain spectrum for the quantum cascade laser of Ref.~\onlinecite{SIR98}
is displayed in Fig.~\ref{FigGainQCL}.
Gain sets in for current densities above 0.8 kA/cm$^2$
and increases with current. The peak gain coefficient is
57 cm$^{-1}$ at 6.5 kA/cm$^2$ and a photon
energy of 130 meV, in excellent agreement
with the findings of Ref.~\onlinecite{SIR98}
(estimated losses divided by confinement factor $63$~cm$^{-1}$ and  
threshold current $J_{\rm th}$=7.2 kA/cm$^2$, lasing at $131$ meV).
The width of the gain spectrum agrees
well with the findings of Ref.~\onlinecite{EIC00}.

\begin{figure}
\includegraphics[width=0.9\columnwidth]{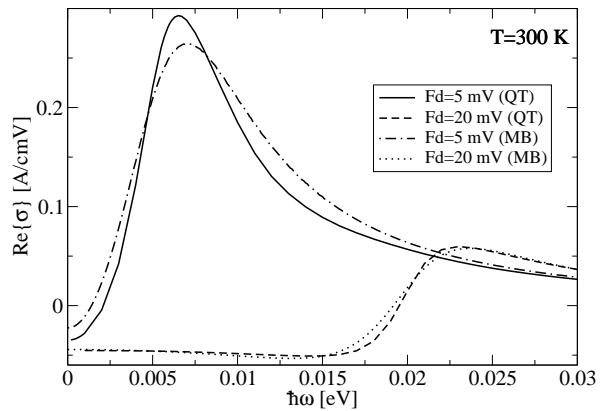}
\caption{Real part of the dynamical conductance for
the superlattice structure of  Ref.~\onlinecite{ALL01}.
Full and dashed line: Result from the quantum transport model
with Coulomb gauge and neglecting self-energy corrections for
$Fd=5$ mV and 20 mV, respectively.
Dotted and dash-dotted line: Corresponding results from the 2-time
miniband model \cite{KTI72}.}
\label{FigSigmaSL}
\end{figure}

Results for the superlattice are given in
Fig.~\ref{FigSigmaSL}. The real part of the conductivity
essentially follows the result from the simple 2-time
miniband model \cite{KTI72}
\begin{equation}
\sigma(\omega)=
\frac{J_{\mathrm MB}(F)}{F}
\frac{1-\omega_B\tau_m\tau_e-\imai \omega\tau_e}
{(\omega_B^2-\omega^2)\tau_m\tau_e+1-\imai \omega(\tau_m+\tau_e)}
\end{equation}
with $\omega_B=eFd/\hbar$.
Thus, this simple model gives, at least for wide minibands
and high temperatures, reliable results.

\begin{figure}
\includegraphics[width=0.9\columnwidth]{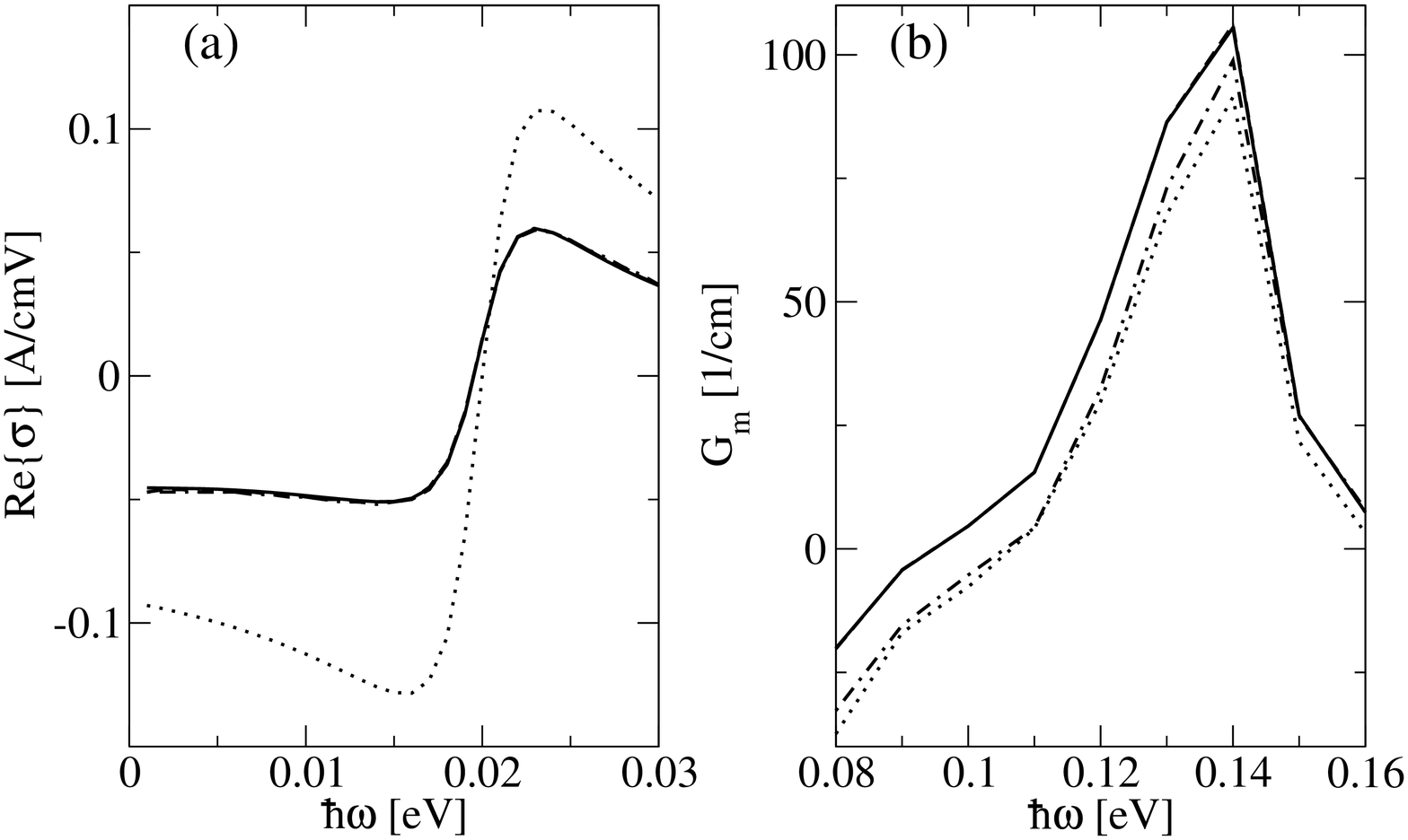}
\caption{Comparison of different choices of the gauge and
 self-energy corrections $\delta\Sigma$
for (a) the superlattice
at $Fd=20$  and (b) the quantum cascade laser at $Fd=260$ mV.
Full line:  Coulomb gauge neglecting $\delta\Sigma$;
Dotted line:  Lorentz gauge neglecting $\delta\Sigma$;
Dashed line: Coulomb gauge with $\delta\Sigma$ 
(falls together with the full line);
Dash-dotted line: Lorentz gauge with $\delta\Sigma$.}
\label{FigGainComp}
\end{figure}

Now we want to study the relevance of the terms 
$\delta \Sigma$ in Eq.~(\ref{EqGlessLinResp}), which have been
neglected so far. Fig.~\ref{FigGainComp} shows
that their inclusion barely changes the result in the Coulomb gauge 
(the full and dashed lines are almost indistinguishable). 
Therefore these self-energy corrections
may be neglected (at least for diagonal self-energies assumed here)
allowing for a significant reduction of the numerical effort.
In contrast, in the case of the Lorentz gauge\footnote{Note that
the translational symmetry of the structure is broken in the
Lorentz gauge}, 
the terms evaluated with
self-energy corrections (dash-dotted line) differ
substantially from the simpler version without these terms
(dotted line).
The difference is particularly strong in the
case of the superlattice, where the diagonal terms in the 
perturbation potential (\ref{EqDeltaU-Lorentz}) are essential for
the transition.
Furthermore, the results of Coulomb gauge and Lorentz gauge
are almost identical if the self-energy corrections are
included. This demonstrates that the
inclusion of these terms is essential to obtain a
consistent theory, which must be independent of the choice
of the gauge.
 
\section{Conclusion} 
A formulation of the theory of nonequilibrium 
Green functions has been given which allows for a quantitative 
description of current-induced gain in semiconductor heterostructures.
Quantitative agreement with the experiment 
for the gain spectrum of a quantum cascade laser was obtained. 
The standard  2-time miniband model for the superlattice
could be verified for a large miniband width and under room temperature
operation. 
The numerical results are not sensitive to the choice of the gauge
if the variations of self-energy terms are taken into account.
These terms are of particular importance for the Lorentz gauge,
while their contribution is small in the Coulomb gauge
in the calculations presented here.

\acknowledgments
Helpful discussions with A.~Knorr, A.-P.~Jauho, S.-C.~Lee, M.~Pereira 
and E.~Sch{\"o}ll as well as financial support by DFG within FOR394
are gratefully acknowledged.

\appendix
\begin{widetext}
\section{Derivation of the equations for the stationary state}
The Green functions are determined
by the Dyson Equation (see, e.g.,  chapter 5 of Ref.~\onlinecite{HAU96})
\begin{multline}
\imai\hbar \pabl{}{t_1}
G^{\text{ret/adv}}_{\alpha_1,\alpha_2}(t_1,t_2)
-\sum_{\beta} U_{\alpha_1,\beta}(t_1)
G^{\text{ret/adv}}_{\beta,\alpha_2}(t_1,t_2)
-\sum_{\beta} \int \frac{\d t}{\hbar}
\Sigma^{\text{ret/adv}}_{\alpha_1,\beta}(t_1,t)
G^{\text{ret/adv}}_{\beta,\alpha_2}(t,t_2)\\
=\hbar\delta(t_1-t_2)\delta_{\alpha_1,\alpha_2}
\label{EqGret-time}
\end{multline}
and the Keldysh relation, see also Appendix \ref{AppRemark}.
\begin{equation}
G^<_{\alpha_1,\alpha_2}(t_1,t_2)=\sum_{\beta,\beta'}
\int \frac{\d t}{\hbar}\int \frac{\d t'}{\hbar}
G^{\text{ret}}_{\alpha_1,\beta}(t_1,t)\Sigma^{<}_{\beta,\beta'}(t,t')
G^{\text{adv}}_{\beta',\alpha_2}(t',t_2)\label{EqKeldysh-time}\, .
\end{equation}

Now we consider the stationary state, where $G(t_1,t_2)=\tilde{G}(t_1-t_2)$.
Eqs.~(\ref{EqGret-time},\ref{EqKeldysh-time}) become:
\begin{gather}
\imai\hbar \pabl{}{t}
\tilde{G}^{\text{ret/adv}}_{\alpha_1,\alpha_2}(t)
-\sum_{\beta} \tilde{U}_{\alpha_1,\beta}
\tilde{G}^{\text{ret/adv}}_{\beta,\alpha_2}(t)
-\sum_{\beta} \int \frac{\d t'}{\hbar}
\tilde{\Sigma}^{\text{ret/adv}}_{\alpha_1,\beta}(t-t')
\tilde{G}^{\text{ret/adv}}_{\beta,\alpha_2}(t')
=\hbar\delta(t)\delta_{\alpha_1,\alpha_2}
\label{EqGret-time-tilde}
\\
\tilde{G}^<_{\alpha_1,\alpha_2}(t)=\sum_{\beta,\beta'}
\int \frac{\d t'}{\hbar}\int \frac{\d t''}{\hbar}
\tilde{G}^{\text{ret}}_{\alpha_1,\beta}(t-t')
\tilde{\Sigma}^{<}_{\beta,\beta'}(t'-t'')
\tilde{G}^{\text{adv}}_{\beta',\alpha_2}(t'')
\label{EqKeldysh-time-tilde}\, .
\end{gather}
and their Fourier-transformation (\ref{EqFourier-tilde})
provides us with Eqs.~(\ref{EqDysonE-tilde},\ref{EqKeldyshE-tilde}).

\section{A remark on Eq.~(5.11) of Ref.~\onlinecite{HAU96}}\label{AppRemark}
Eq.~(\ref{EqKeldysh-time}) is a particular solution
of the inhomogeneous
differential equations (5.3) of Ref.~\onlinecite{HAU96}, 
which are given more explicitly in Eqs.~(106-107) of 
Ref.~\onlinecite{WAC02} in the notation used here.
The Keldysh relation (5.11) derived in Ref.~\onlinecite{HAU96} contains
an additional term:
\begin{multline}
{\bf G}_{\textrm{add}}^<(t_1,t_2)
=\int \frac{\d t_b}{\hbar}
\int \frac{\d t_c}{\hbar}
\left[
\hbar\delta(t_1-t_b){\bf 1}+
{\bf G}^{\mathrm{ret}}(t_1,t_b){\bf U}(t_b)
+\int \frac{\d t_a}{\hbar}
{\bf G}^{\mathrm{ret}}(t_1,t_a)
\bm{\Sigma}^{\mathrm{ret}}(t_a,t_b)\right]\\
\times{\bf G}_0^<(t_b,t_c)
\left[\hbar\delta(t_c-t_2){\bf 1}
+{\bf U}(t_c){\bf G}^{\mathrm{adv}}(t_c,t_2)
+\int \frac{\d t_d}{\hbar}\bm{\Sigma}^{\mathrm{ret}}(t_c,t_d)
{\bf G}^{\mathrm{adv}}(t_d,t_2)\right]
\end{multline}
This term constitutes a homogeneous solution
of the inhomogeneous differential equations (for fixed $\Sigma$ and 
$G^{\text{ret/adv}}$) which
guarantees that $G^<\to G^<_0$ for $t_1,t_2\to -\infty$.
This satisfies a general assumption underlying the
perturbation expansion of the $S$-matrix.
In the following we show that $G^<_{\textrm{add}}$ 
vanishes for finite times $t_1,t_2$.

Using Eqs.~(108,109) of Ref.~\onlinecite{WAC02} one obtains
\begin{equation}
\begin{split}
{\bf G}_{\textrm{add}}^<(t_1,t_2)
=&\int \frac{\d t_a}{\hbar}
\int \frac{\d t_c}{\hbar}
{\bf G}^{\mathrm{ret}}(t_1,t_a)
\left(-\imai \hbar\pabl{^{\mathrm{left}}}{t_a}
-{\bf E}\right)
{\bf G}_0^<(t_a,t_c)
\left(\imai \hbar\pabl{}{t_c}-{\bf E}\right)
{\bf G}^{\mathrm{adv}}(t_c,t_2)\\
=&\int \frac{\d t_a}{\hbar}
\int \frac{\d t_c}{\hbar}
{\bf G}^{\mathrm{ret}}(t_1,t_a)
\left(\imai \hbar\pabl{}{t_a}
-{\bf E}\right)
{\bf G}_0^<(t_a,t_c)
\left(\imai \hbar\pabl{}{t_c}-{\bf E}\right)
{\bf G}^{\mathrm{adv}}(t_c,t_2)\\
&-\imai\left[\int \frac{\d t_c}{\hbar}
{\bf G}^{\mathrm{ret}}(t_1,t_a)
{\bf G}_0^<(t_a,t_c)
\left(\imai \hbar\pabl{}{t_c}-{\bf E}\right)
{\bf G}^{\mathrm{adv}}(t_c,t_2)\right]_{t_a=-\infty}^{t_a=\infty}\, ,
\end{split}
\end{equation}
where $\partial^{\mathrm{left}}$ means that the derivative operates
to the left. In the second line, partial integration has been used.
As $\left(\imai \hbar\pabl{}{t_a}-{\bf E}\right)
{\bf G}_0^<(t_a,t_c)=0$, the first term vanishes. Furthermore,
at the upper bound ${\bf G}^{\mathrm{ret}}(t_1,\infty)=0$ holds for
finite times $t_1$. Applying the same manipulations for $t_c$, we
find
\begin{equation}
\begin{split}
{\bf G}_{\textrm{add}}^<(t_1,t_2)
=&\imai\int \frac{\d t_c}{\hbar}
{\bf G}^{\textrm{ret}}(t_1,-\infty)
{\bf G}_0^<(-\infty,t_c)
\left(\imai \hbar\pabl{}{t_c}-{\bf E}\right)
{\bf G}^{\textrm{adv}}(t_c,t_2)\\
=&{\bf G}^{\textrm{ret}}(t_1,-\infty)
{\bf G}_0^<(-\infty,-\infty)
{\bf G}^{\textrm{adv}}(-\infty,t_2)\, .
\end{split}
\end{equation}
For a real system the retarded and advanced 
Green functions decay for large time differences
and we find ${\bf G}_{\textrm{add}}^<(t_1,t_2)\to 0$ for finite
$t_1,t_2$. Therefore this additional term is spurious.

A second, more physical argument relates to the
idea, that  the stationary state of a real physical system should 
not depend on the initial conditions used in a formal theory.
Except for specific situations exhibiting a hysteresis, scattering
processes will install a stationary state which is
independent of initial conditions. Therefore all additional
homogeneous solutions should reflect transient effects,
which vanish if the initial state is chosen at $t_1=-\infty$ and
$t_2=-\infty$.

\section{Derivation of Eqs.~(\ref{EqGretLinResp},\ref{EqGlessLinResp})}
\label{AppDer}
Setting $U(t)=\tilde{U}+\delta U(t)$, 
$G(t_1,t_2)=\tilde{G}(t_1-t_2)+\delta G(t_1,t_2)$  and
$\Sigma(t_1,t_2)=\tilde{\Sigma}(t_1-t_2)+\delta \Sigma(t_1,t_2)$ in
Eqs.~(\ref{EqGret-time},\ref{EqKeldysh-time}) gives in
linear order $\delta U$:
\begin{multline}
\imai\hbar \pabl{}{t_1}
\delta G^{\text{ret}}_{\alpha_1,\alpha_2}(t_1,t_2)
-\sum_{\beta} \tilde{U}_{\alpha_1,\beta}
\delta G^{\text{ret}}_{\beta,\alpha_2}(t_1,t_2)
-\sum_{\beta} \int \frac{\d t}{\hbar}
\tilde{\Sigma}^{\text{ret}}_{\alpha_1,\beta}(t_1-t)
\delta G^{\text{ret}}_{\beta,\alpha_2}(t,t_2)
=\\
\sum_{\beta} \delta U_{\alpha_1,\beta}(t_1)
\tilde{G}^{\text{ret}}_{\beta,\alpha_2}(t_1-t_2)
+\sum_{\beta} \int \frac{\d t}{\hbar}
\delta\Sigma^{\text{ret}}_{\alpha_1,\beta}(t_1,t)
\tilde{G}^{\text{ret}}_{\beta,\alpha_2}(t-t_2)
\end{multline}
This constitutes an inhomogeneous differential equation,
where the homogeneous part (left-hand side) is identical with the defining
equation (\ref{EqGret-time-tilde}) for $\tilde{G}^{\text{ret}}$.
Thus  $\tilde{G}^{\text{ret}}$ is the 
corresponding Green function and the solution is given by:
\begin{equation}
\delta {\bf G}^{\text{ret}}(t_1,t_2)
=\int \frac{\d t}{\hbar}\int 
\frac{\d t'}{\hbar}
\tilde{\bf G}^{\text{ret}}(t_1-t)\Big[\delta {\bf U}(t)\hbar\delta(t-t')
+\delta\bm{\Sigma}^{\text{ret}}(t,t')\Big]       
\tilde{\bf G}^{\text{ret}}(t'-t_2)
\end{equation}
and Fourier transformation using Eq.~(\ref{EqFourier}) gives 
Eq.~(\ref{EqGretLinResp}) after some algebra.
$\delta G^{\text{adv}}$ is evaluated in the same way with 
ret replaced by adv.

The linearization of the Keldysh relation (\ref{EqKeldysh-time})
provides us with
\begin{multline}
\delta {\bf G}^{<}(t_1,t_2)=\int \frac{\d t}{\hbar}\int \frac{\d t'}
{\hbar}
\Big[\delta{\bf G}^{\text{ret}}(t_1,t)\tilde{\bm{\Sigma}}^{<}(t-t')
\tilde{\bf G}^{\text{adv}}(t'-t_2)
+\tilde{\bf G}^{\text{ret}}(t_1-t)\delta\bm{\Sigma}^{<}(t,t')
\tilde{\bf G}^{\text{adv}}(t'-t_2)\\
+\tilde{\bf G}^{\text{ret}}(t_1-t)
\tilde{\bm{\Sigma}}^{<}(t-t')\delta{\bf G}^{\text{adv}}(t',t_2)\Big]
\end{multline}
Fourier transformation  gives 
Eq.~(\ref{EqGlessLinResp}) after some algebra.

\section{Choice of gauges}\label{AppGauge}
We study the linear response to an external radiation field
propagating in $y$-direction
\begin{equation}
\vec{F}(\vec{r},t)=\int \frac{\d \omega}{2\pi}
F(\omega)\vec{e}_z\e^{\imai k(\omega) y-\imai\omega t}
\qquad
\vec{B}(\vec{r},t)=\int \frac{\d \omega}{2\pi}
\frac{k(\omega)}{\omega}F(\omega)\vec{e}_x
\e^{\imai k(\omega) y-\imai\omega t}
\, .
\end{equation}
Thus we have two reasonable choices of the electromagnetic potentials
\cite{JAC98a}:
\begin{alignat}{3}
&\mbox{Coulomb gauge}& \quad
\vec{A}(\vec{r},t)&=\int \frac{\d \omega}{2\pi}
\frac{F(\omega)}{\imai \omega}\vec{e}_z
\e^{\imai k(\omega) y-\imai\omega t}
& \quad \phi(\vec{r},t)&=0\\
&\mbox{Lorentz gauge}& \quad
\vec{A}(\vec{r},t)&=-\int \frac{\d \omega}{2\pi}
\frac{k(\omega)F(\omega)z}{\omega}
\vec{e}_y\e^{\imai k(\omega) y-\imai\omega t}
& \quad \phi(\vec{r},t)&=-F(\omega)z\e^{\imai k(\omega)y-\imai\omega t}
\end{alignat}
In the spirit of a dipole approximation
(i.e., assuming that the wavelength 
is large compared to the size of the active region)
the terms containing
$k$ are neglected, providing the expressions
(\ref{EqDeltaU-Coulomb},\ref{EqDeltaU-Lorentz})
in the perturbation Hamiltonian, where quadratic terms
$\propto F(\omega)^2$ are neglected.
\end{widetext}


\end{document}